# Competitive adsorption of surfactants and polymers on colloids by means of mesoscopic simulations

Armando Gama Goicochea

Departamento de Ciencias Naturales, DCNI, Universidad Autónoma Metropolitana Unidad Cuajimalpa, Av. Pedro Antonio de los Santos 84, México, D. F. 11850, Mexico.

E-mail: agama@alumni.stanford.edu

## Abstract

The study of competitive and cooperative adsorption of functionalized molecules such as polymers, rheology modifiers and surfactant molecules on colloidal particles immersed in a solvent is undertaken using coarse – grained, dissipative particle dynamics simulations. The results show that a complex picture emerges from the simulations, one where dispersants and surfactants adsorb cooperatively up to certain concentrations, on colloidal particles, but as the surfactant concentration increases it leads to dispersant desorption. The presence of rheology modifying agents in the colloidal dispersion adds complexity through the association of surfactant micelles to hydrophobic sites of these agents. Analysis of the simulation results reported here point clearly to the self-association of the hydrophobic sites along the different polymer molecules as the mechanism responsible for their competitive and cooperative adsorption.

## 1 Introduction

Polymer adsorption is crucial for the performance in modern applications of complex fluids, such as in stimuli – responsive systems, biological membranes, and consumer goods such as paints, cosmetics or food products. In particular, polymer adsorption on pigments surfaces remains a popular mechanism to stabilize architectural paints [Napper, 1983]. There are other types of polymeric molecules that can also be adsorbed on particles, such as surfactants and rheology modifying agents. These functionalized molecules have usually different lengths and interact not only with each other and the solvent, but also with themselves.

The characterization of polymer and surfactant adsorption is usually carried out through measurements of adsorption isotherms, which yield directly information about the optimal amount of polymer needed to achieve stability [Kronberg, 2001]. However, the simultaneous presence of more than one type of polymers in the dispersion can give rise to a complex combination of competition and synergy between polymer molecules, which leads to competitive adsorption isotherms. These types of experiments are laborious and time consuming, taking up to several weeks to complete. One attractive alternative is the use of molecular modeling using appropriately adapted algorithms for relatively complex fluids, which can then be solved highly accurately using modern computers.

This work is devoted to the presentation of coarse – grained computer simulations for the prediction and understanding of competitive adsorption isotherms of polymers and surfactants on colloidal particles. It is argued that the mesoscopic reach of the simulations carried out is especially important to obtain results that are directly comparable with scales probed with experiments on soft matter systems. This study, which is the first of its kind to the best of the author's knowledge, is a useful representation of architectural paints and coatings, as well as of other complex fluids of current academic and industrial interest.

## 2 Models, Methods and Systems

The force model used in the simulations presented here is a mesoscopic, coarse – grained method known as dissipative particle dynamics (DPD) [Hoogerbrugge and Koelman, 1982]. It involves central, pairwise forces between DPD beads, which are not physical particles but rather momentum – carrying sections of the fluid. There are three types of forces in the DPD model: a conservative force ($F_C$), which determines the local hydrostatic pressure; a dissipative force ($F_D$), that represents the local viscosity of the fluid, and a random force ($F_R$), constituted by the Brownian motion of DPD beads. The latter two forces exactly balance each other by construction, as a result of the fluctuation – dissipation theorem (Groot and Warren, 1997); this feature is the essence of the DPD model. The functional dependence of the forces is not specified by the DPD model, but they are usually chosen as simple as possible; the most employed ones are repulsive, linearly decaying (for $F_C$) and short ranged. The structure of the DPD model, as well as some of its strengths and weaknesses are well known, and the reader is referred to recent reviews, like the one by Murtola et al. [Murtola et al. 2009] for details.

The systems studied are constituted by the polymeric molecules of different functionality (surfactants, dispersant polymers, rheology modifiers), the solvent (water), and the colloidal particles (pigments, fillers). The latter are typically much larger than the rest, so one can consider them as flat surfaces fixed in space, and then invest the computational cost on solving the motion of the rest of the particles. Although these polymeric molecules share the characteristic that they are amphiphilic in nature, they are usually distinguished by the role they play. Hence, surfactants are typically short molecules whose only purpose is to reduce the surface or interfacial tension. Dispersant polymers are longer and they are used to adsorb on colloids and keep them apart, hence their name. Rheology modifiers are large polymeric molecules, generally made of units of different chemical nature, with hydrophobic and hydrophilic parts. Their function is to modify the viscosity of the fluid in which they are dissolved. The polymeric molecules are constructed as DPD beads joined by freely rotating harmonic springs, and can be linear or branched; the solvent is represented by single beads and for the surfaces, an effectively exact DPD wall force is used, given by a repulsive, short range polynomial [Gama Goicochea and Alarcón, 2011]. For the surfactant, a non – ionic, linear, 14 – bead polymer was used as a model for a nonylphenol etoxylate surfactant. The dispersant was modeled as a 48 – DPD bead linear polymer, to represent a hydrophobic dispersant made of maleic anhydride and styrene. As for the rheology-modifying agent, I used a hydrophobically modified alkali-swellable emulsion (HASE) polymer, which is represented by 60 DPD units. In regards to the conservative DPD force interaction parameters, they have been chosen following the standard procedure (Groot and Warren, 1997), beginning with the isothermal compressibility of water at room temperature to choose the equal – particle interaction. For different particles interaction, the Flory – Huggins parameter is used based on the chemical composition of each DPD bead. As for the choice of wall – DPD particle force, it has been chosen by fitting the interfacial tension values of the confined fluid with the wall – particle value. The interaction parameters as well as the specific bead sequence shall be omitted for brevity but may be consulted in reference [Gama Goicochea, 2013], along with all simulation details.

Adsorption experiments are generally performed following a route in which the polymers to be adsorbed are added to the system and the measurements are performed when chemical, thermal and mechanical equilibrium is achieved. To properly reproduce those conditions, the simulations are carried out in the grand canonical thermodynamic ensemble, where the chemical potential, temperature and volume are kept constant as the polymer concentration is increased. The DPD method has been adapted to the grand canonical ensemble (constant chemical potential, volume and temperature) to obtain the competitive adsorption isotherms presented here. The procedure is the following: the volume of the simulation box is fixed ($L_x=L_y=7$; $L_z=14$ DPD dimensionless units), then a fixed number of one type of additives, say, dispersant polymers is added to it, along with a fixed number of rheology modifying agents. Then, the adsorption is monitored by adding molecules of, for example, surfactants to the box and performing the simulations until equilibrium has been achieved, while the temperature, volume and chemical potential are kept fixed. The chemical potential is fixed through the exchange of solvent particles with the virtual bulk. In the simulations, this chemical potential was fixed at $\mu=37.7$ units so that the total average density in the simulation box was nearly equal to 3. In doing so, one ensures that the equation of motion of the DPD fluid is invariant under changes of the interaction parameters (Groot and Warren, 1997). Full details of the DPD algorithm adapted to the grand canonical ensemble, as well as simulation details such as the integration algorithm, time step, simulation length, etc., can be found in [Gama Goicochea, 2007].

## 3 Results

Let us first illustrate the capabilities of the DPD method by presenting the association of a surfactant molecule with a single HASE (rheology modifier) molecule. The system consists of 60 surfactant molecules, in addition to the HASE molecule, in solution with solvent molecules. No colloidal particles were present therefore periodic boundary conditions were used in all directions. All molecules positions are chosen at random initially and are allowed to evolve, subjected to the DPD forces. Figure 1 shows the final configuration obtained after equilibrium was reached.

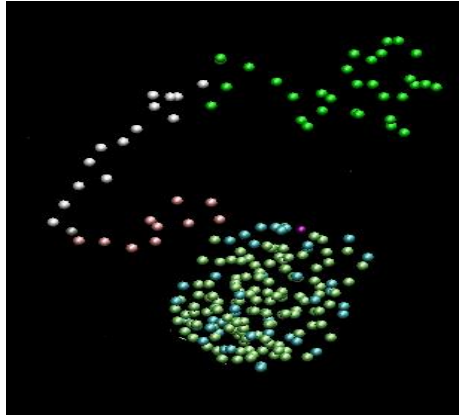

**Fig. 1.** Equilibrium configuration of a single linear molecule of a rheology-modifying agent (HASE) with a surfactant micelle formed at one of its hydrophobic sites. The colors represent the different chemical characteristics of the molecules (see Gama Goicochea (2013) to see the exact chemical composition and DPD mapping). The hydrophilic parts of the HASE and surfactant molecules, as well as the solvent molecules are omitted for clarity.

As suggested by Fig. 1, HASE molecules modify the rheology of fluids by promoting the formation of surfactant micelles on specific hydrophobic sites on the HASE backbone. Self – association, and association between different HASE molecules can then be modulated through the judicious choice of surfactants, which in turn will modify the rheology of the fluid. This obviously follows from Figure 1: when many HASE molecules are present in a solution with surfactants, they shall tend to associate as shown in Figure 1 and therefore an association between HASE molecules will be unavoidable due to the steric interactions between those complex molecular conglomerates. Figure 1 represents a textbook example [Glass, 2000] of the mechanism through which these types of molecules are thought to associate, but here it has been shown to emerge from molecular modeling.

I shall now proceed to the presentation of the adsorption isotherms, of which 2 different types were calculated. One, where the dispersant polymer concentration was fixed while the surfactant concentration was increased, and one where it was the surfactant concentration what was kept fixed while the dispersant concentration was varied. The purpose of carrying out the adsorption isotherms through these two routes is deciding which procedure leads to the optimal dispersion conditions. The fluid in all cases is confined by two different types of surfaces: one is a metal oxide, $TiO_2$, and the other is a silicate-based colloid with almost negligible interactions with the polymers involved in this study, whose only purpose is that of occupying space, hence its name "filler". The parameters of interaction between these surfaces and the DPD fluid have been tested and have been successfully used before, see Gama Goicochea (2007) and Gama Goicochea (2013).

In the left of Figure 2 I show the adsorption isotherm of the surfactant when the dispersant and the thickener (HASE) concentrations are fixed. It may appear that the surfactant adsorption is hardly influenced by the presence of the other types of polymers in the dispersion, for the saturation concentration of the surfactant remains almost constant. However, when the isotherm of the surfactant is obtained, at fixed dispersant concentration (without rheology modifiers), which is shown in the inset, the number of

Competitive adsorption of surfactants and polymers on colloids adsorbed surfactant molecules is found to increase slowly with added surfactant concentration. Hence, there is clearly an interplay between the surfactant and the dispersant, which enhances the adsorption of the surfactant by the thickener, cooperatively. While the surfactant adsorption is greatly influenced by the thickener and the filler, the dispersant is not. This conclusion is obtained from the right panel in Figure 2.

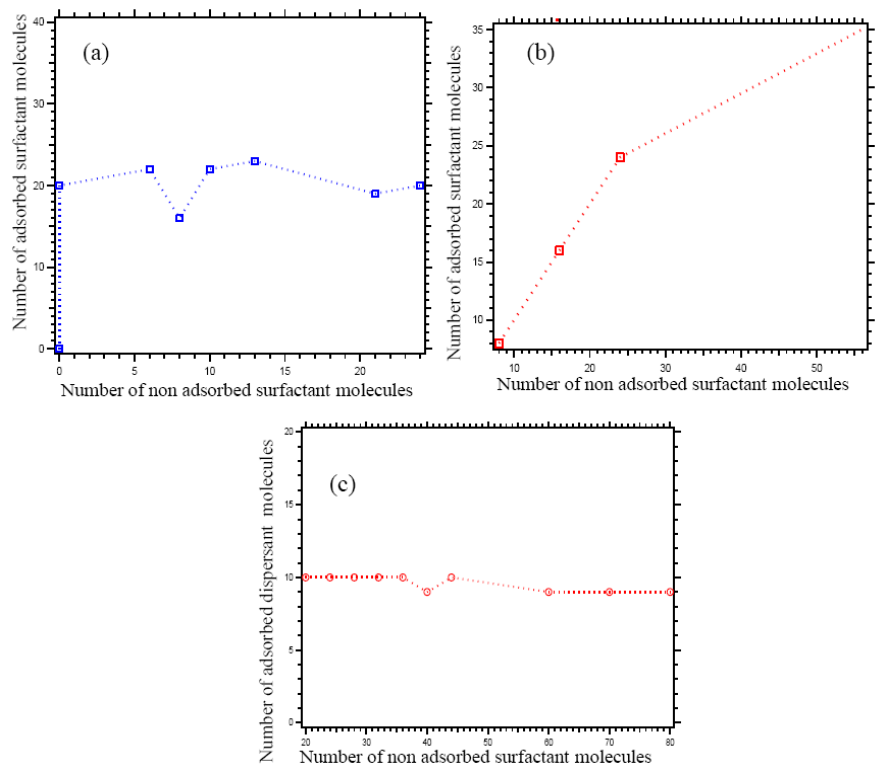

**Fig. 2.** Adsorption surfactant isotherm obtained for (a) fixed dispersant concentration (10 dispersant molecules, with the number of surfactant molecules varying from 20 up to 80) and (c) dispersant adsorption isotherm at a fixed surfactant concentration (10 surfactant molecules, with the number of dispersant molecules ranging from 6 up to 40). Figure 2(b) shows the single (non competitive) isotherm for the surfactant alone. For cases (a) and (c) the system contains 6 HASE molecules and is confined by flat walls representing $TiO_2$ and a filler (silicate-based colloid) surfaces.

The isotherm on the right in Figure 2, which corresponds to that of the dispersant at fixed surfactant and rheology modifier concentrations, is not at all perturbed by these additives. When the adsorption isotherm for the dispersant only was calculated (not shown, for brevity), the same trend was obtained, namely, a constant saturation concentration, as shown on the right panel in Figure 2. Therefore, the adsorption mechanisms that take place even if the components of the colloidal dispersion are the same, can change radically depending on the variable of control.

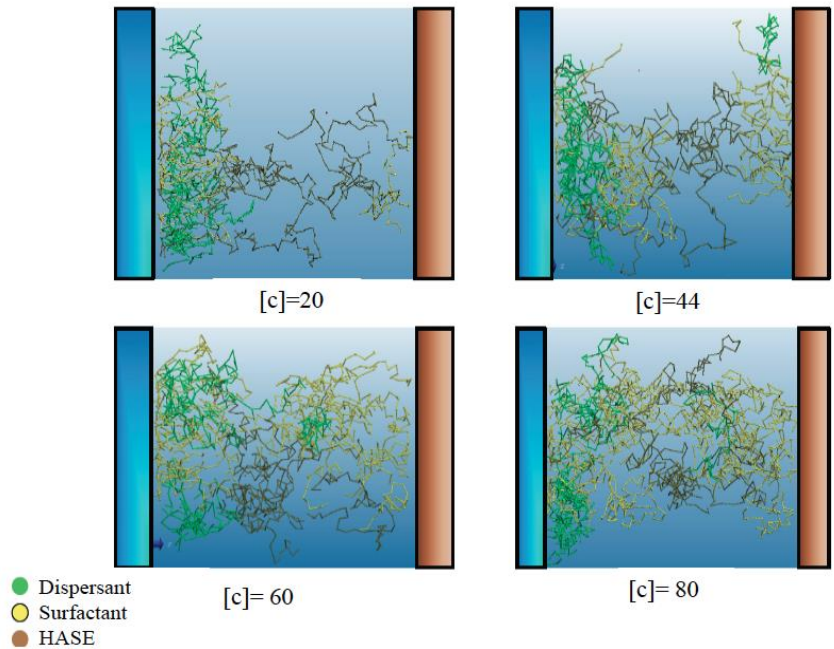

**Fig. 3.** Configuration of the dispersant (green), surfactant (yellow) and rheology modifier (brown) molecules as the surfactant concentration is increased, from 20 up to 80 molecules. For all cases the system contains 10 dispersant molecules and 6 HASE molecules and is confined by flat walls representing $TiO_2$ (left) and filler (right) surfaces. The solvent has been removed for clarity.

Competitive adsorption of surfactants and polymers on colloids

A clear image of the evolution of the adsorption process which may not be appreciated from the isotherms alone can perhaps be better gained from inspection of Figure 3. In it I show snapshots obtained from the DPD simulations, after equilibrium was reached. At the lowest surfactant concentration ([c]=20) all the dispersant is adsorbed on the $TiO_2$ surface, with the thickener almost completely extended and the surfactant associated with the dispersant. As the surfactant concentration is increased to [c]=40 molecules, some of the dispersant molecules were desorbed and even migrated to the filler substrate, on the right. At the largest surfactant concentrations, the dispersant got even more desorbed, with the surfactant replacing it at the adsorption sites, on both substrates. The thickener shows self association (see the middle of the simulation box) and the dispersant prefers to associate with the surfactant and the thickener rather than remain adsorbed. Evidently, at low concentrations the surfactant promotes the adsorption of the dispersant, i.e., they behave synergistically, whereas at large surfactant concentrations the opposite happens.

Precisely this type of behavior has been observed in experiments of competitive adsorption carried out with polymers and cationic, anionic and nonionic surfactants [Karlson et al., 2008] where the authors found that at low surfactant concentration, the polymer (which plays the role of the dispersant) remained adsorbed (on polystyrene and silica particles) while the surfactant formed micelles. As the concentration of the surfactant was increased, and if the polymer and the surfactant attract, they form complexes that can be desorbed. If one of them, be it the surfactant or the polymer has higher affinity for the surface, it will replace the other on the particle surface. The conclusions derived from the experimental model, water – based paint designed by Karlson and co workers are fully supported by the results of the simulations presented in this work.

The simulations presented here give additional insight into why the phenomena presented in Figures 2 and 3 occur. Figure 4 shows the density profiles of the hydrophobic sections of all three types of polymers in the colloidal dispersion.

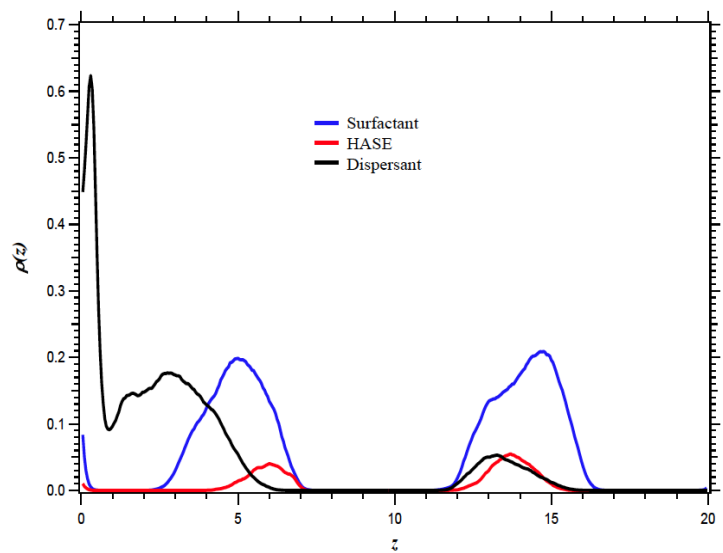

**Fig. 4.** Density profiles of the hydrophobic sections of the surfactant (blue), thickener (red) and dispersant (black). The pigment surface is the one on the left and the filler surface is on the right.

The density profiles shown in Figure 4 show that the polymers associate because of the affinity of their hydrophobic sections, as clearly indicated by the maxima around $z=5$ and $z=15$ (dimensionless units). Although most of the dispersant remains adsorbed on the $TiO_2$ surface (on the left), some of it desorbed and formed a complex associated structure with the surfactant and the rheology modifier close to the pigment. Additionally, the surfactant formed a micelle around the hydrophobic sites of the thickener, and some dispersant molecules were completely desorbed and associated with the surfactant micelle, as shown by the structure form around $z=15$. Obviously this behavior arises from basic molecular hydrophobic interactions due to the structure and characteristics of the polymers modeled in these simulations.

## 4 Conclusions

The complex mechanisms that give rise to competitive and cooperative adsorption of polymers with different functional groups in a colloidal dispersion were studied for the first time, using mesoscopic, DPD computer simulations. Two different colloidal particles were included: a pigment ($TiO_2$) and a silicate-based filler. The surfactant, dispersant and rheology modifying polymers were found to associate cooperatively at low surfactant concentration, promoting the full adsorption of the dispersant which, in turn, leads to a more stable paint. This is the result of the affinity that the hydrophobic groups present in all three types of molecules have. However, as the surfactant concentration is increased, the same affinity of the hydrophobic groups makes it energetically and entropically more advantageous for some of the dispersant molecules to be desorbed, forming micelles with the thickener that eventually lead to a less stable dispersion. It was argued that these conclusions are fully supported by recent experiments on model paints. This work is expected to be useful not only to formulators and expert designers of modern water – based paints and coatings, but also to those studying smart materials and biological membranes.

## 5 Acknowledgements

The author is indebted to the following individuals for enlightening discussions: F. Alarcón, M. Briseño, N. López, A. Ortega, H. Ortega, E. Pérez, and F. Zaldo. This work was sponsored in its initial phase by the Centro de Investigación en Polímeros (Grupo Comex), and afterward by PROMEP through project 47310286-912025.